\documentclass[preprint,aps,pra,showpacs,floatfix]{revtex4}
\usepackage{graphicx}
\usepackage{times}
\usepackage{nicefrac}
\usepackage{amsmath}
\usepackage{amsfonts}
\usepackage{amssymb}
\usepackage{amsthm}
\usepackage{epsf}
\usepackage{bm}
\usepackage{bbm}
\usepackage{times}

\usepackage{dcolumn}
\newcolumntype{.}{D{x}{}{-1}}

\newcommand{\balpha}{{\mbox{\boldmath$\alpha$}}}

\newcommand{\bsigma}{{\mbox{\boldmath$\sigma$}}}
\newcommand{\bSigma}{{\mbox{\boldmath$\Sigma$}}}
\newcommand{\be}{\begin{eqnarray}}
\newcommand{\ee}{\end{eqnarray}}

\newcommand{\az}{\alpha Z}
\newcommand{\bfA}{{\bf A}}

\newcommand{\bfp}{{\bf p}}
\newcommand{\bfr}{{\bf r}}

\newcommand{\calH}{{\cal H}}
\newcommand{\calE}{{\cal E}}

\newcommand{\Dmatrix}[4]{
        \left(
        \begin{array}{cc}
        #1  & #2   \\
        #3  & #4   \\
        \end{array}
        \right)
        }

\newcommand{\mub}{\mu_0}
\newcommand{\vmagn}{V^{\rm magn}}
\newcommand{\gdirac}{g_{\rm D}}
\newcommand{\dgint}{\Delta g_{\rm int}}
\newcommand{\dgqed}{\Delta g_{\rm QED}}
\newcommand{\dgsqed}{\Delta g_{\rm SQED}}
\newcommand{\dgrec}{\Delta g_{\rm rec}}
\newcommand{\dgns}{\Delta g_{\rm NS}}
\newcommand{\Hint}{H^{\rm int}}

\newcommand{\HC}{H^{\rm C}}
\newcommand{\HB}{H^{\rm B}}
\newcommand{\dgdf}{\Delta g_{\rm int}^{\rm CI-DF}}
\newcommand{\dgho}{\Delta g_{\rm int}^{(2+)}}
\newcommand{\dgexact}{\Delta g_{\rm int}^{(1)}}

\newcommand{\dgse}{\Delta g_{\rm SE}}
\newcommand{\dgvp}{\Delta g_{\rm VP}}

\newcommand{\dgvpe}{\Delta g_{\rm VP}^{\rm e}}
\newcommand{\dgvpeu}{\Delta g_{\rm VP}^{\rm e,U}}
\newcommand{\dgvpewk}{\Delta g_{\rm VP}^{\rm e,WK}}
\newcommand{\dgvpm}{\Delta g_{\rm VP}^{\rm m}}

\newcommand{\dgsqedz}{\Delta g_{\rm SQED}^{(1)}}
\newcommand{\dgsqedzz}{\Delta g_{\rm SQED}^{(2+)}}

\newcommand{\gfree}{g_{\rm free}}
\newcommand{\hrad}{h^{\rm rad}}

\newcommand{\Hrada}{H^{\rm rad}_1}
\newcommand{\Hradb}{H^{\rm rad}_2}
\newcommand{\Hradc}{H^{\rm rad}_3}

\begin{document}

\title{Relativistic and QED corrections to the $\bm{g}$ factor of Li-like ions}
\author{D.~A.~Glazov,$^{1,2}$ V.~M.~Shabaev,$^{1,3}$ I.~I.~Tupitsyn,$^{1}$
A.~V.~Volotka,$^{1,3}$ V.~A.~Yerokhin,$^{1,2,4}$
G.~Plunien,$^{3}$ and G.~Soff$^{3}$}

\affiliation{
$^1$
Department of Physics, St. Petersburg State University, Oulianovskaya 1,
Petrodvorets, St. Petersburg 198504, Russia\\
$^2$
Max-Planck-Institut f\"ur Physik Komplexer Systeme, N\"othnitzer Str. 38,
D-01187 Dresden, Germany\\
$^3$ Institut f\"ur Theoretische Physik, TU Dresden, Mommsenstra{\ss}e 13,
D-01062 Dresden, Germany \\
$^4$ Center for Advanced Studies, St. Petersburg State Polytechnical
University, Polytekhnicheskaya 29, St. Petersburg 195251, Russia\\
}

\begin{abstract}
Calculations of various corrections to the $g$ factor of Li-like ions are
presented, which result in a significant improvement of the theoretical
accuracy in the region $Z=6\,$--$\,92$. The configuration-interaction
Dirac-Fock method is employed for the evaluation of
the interelectronic-interaction correction of order $1/Z^2$ and higher.
This correction is combined with the $1/Z$ interelectronic-interaction term 
derived within a rigorous QED approach.
The one-electron QED corrections of first in $\alpha$
are calculated to
all orders in the parameter $\az$. The screening of QED corrections is taken
into account to the leading orders in $\az$ and $1/Z$.
\end{abstract}


\maketitle

%
\section{Introduction}

Recent high-precision measurements of $g$ factors of H-like carbon
\cite{her00,haf00} and oxygen \cite{ver04} have provided a possibility for
studying binding QED effects in an external magnetic field in these systems.
The experiments were performed on single H-like ions confined in a Penning
ion trap at low temperatures. A relative accuracy of $5\times 10^{-10}$ was
obtained in Ref.~\cite{haf00} for the ratio of the electronic Larmor
precession frequency $\omega_{L}$ and the ion cyclotron frequency $\omega_{
c}$, which is connected with the $g$-factor value by
\begin{equation}
\frac{\omega_{L}}{\omega_{c}} = \frac{g}{2}\, \frac{|e|}{q}\, \frac{m_{\rm
ion}}{m}\,, \label{1}
\end{equation}
where  $e$ is the elementary charge unit, $q$ is the charge of the ion,
$m_{\rm ion}$ is the ion mass, and $m$ is the electron mass. The experimental
results are shown to be sensitive to one- and two-loop binding QED effects
and to the nuclear-recoil corrections. Further progress is anticipated from
the experimental side, as well as an extension of measurements to the ions
with more than one electron.

New perspectives for testing QED effects in $g$ factors of highly charged
ions motivated numerous theoretical investigations on this subject during the
last years. We mention here numerical calculations of the one-loop
self-energy correction to all orders in $\az$
\cite{blu97,per97pra,bei00pra,yer02prl,yer04pra}, numerical
\cite{per97pra,bei00pra} and analytical
\cite{karshenboim:00:pla,kar01a,kar02plb} evaluations of the
one-loop vacuum-polarization contribution, analytical results for the
nuclear-size correction \cite{karshenboim:00:pla,gla02pla}, and calculations
of the nuclear-recoil effects \cite{sha02pra:rec,yel01,sha02,martynenko:01}
and two-loop binding QED corrections
\cite{cza01pra,kar:01:hydr,pachucki:04}. As a result of these studies,
the present theoretical accuracy of the $g$-factor values is several times
better than that of the experimental results. An important consequence of the
investigations of the $g$ factor is the possibility to determine the electron
mass from Eq.~(\ref{1}) by combining the theoretical $g$-factor value and the
experimental result for $\omega_L/\omega_c$. We note that the 1998 CODATA
value for the electron mass \cite{moh00rmp} has an error of $2\times10^{-9}$,
which is 4 times larger than the experimental uncertainty of the result for
carbon \cite{haf00} and 3 times larger than the one for oxygen \cite{ver04}.
A new determination of the electron mass presented in
Refs.~\cite{yel01,bei02prl,yer02prl,ver04} provided an improvement of
the accuracy of the electron mass  by a factor of 4. As a result, the 2002
CODATA value for the electron mass \cite{CODATA02} is derived mainly from
the $g$ factor of H-like ions. It is expected that in the future an extension
of experimental investigations towards higher-$Z$ ions could lead also to an
independent determination of the fine structure constant $\alpha$
\cite{kar:01:hydr,werth:01:hydgr}.

The accuracy of theoretical values for the $g$ factor of high-$Z$ H-like ions
is presently limited by nuclear effects \cite{sha02,nef02prl}. The uncertainty
introduced by them in the high-$Z$ region is comparable with the binding
QED correction of second order in $\alpha$. Since the nuclear effects
do not allow an accurate theoretical description at present, this puts
a serious obstacle on the way to improvement of theoretical predictions and to
an identification of two-loop QED corrections in future experiments. However,
it was recently shown \cite{sha02pra,sha03nim} that the uncertainty due to the nuclear
effects can be significantly reduced in a specific combination of the $g$
factors of H- and Li-like ions with the same nucleus,
\begin{equation}
g ^{\prime} = g_{(1s)^2\,2s}-\xi\, g_{1s}\,,
\end{equation}
where $g_{1s}$ and $g_{(1s)^2\,2s}$ are the $g$ factors of H- and Li-like
ions, respectively, and the parameter $\xi$ is calculated numerically, as
explained in Ref.~\cite{sha02pra}. Numerical calculations show that the
influence of the nuclear effects on the difference $g^{\prime}$ is by two
orders of magnitude smaller than that on the $g$-factor values $g_{1s}$ and
$g_{(1s)^2\,2s}$ separately. Therefore, the specific difference $g^{\prime}$
can be in principle studied up to much higher accuracy than the $g$ factor of
an H-like or Li-like ion. In order to realize this project, one need to
perform theoretical and experimental investigations of the $g$ factor of
Li-like ions with an accuracy comparable to that for H-like ions.

Extending theoretical description from an H-like to a Li-like ion, one
encounters a serious complication due to the presence of additional
electrons. Moreover, all contributions to the $g-2$ value for an $ns$ state
are of pure relativistic and QED origin, which makes the treatment of the
electron correlation much more intricate than, e.g., in calculating the
binding energies. A number of relativistic calculations of the $g$ factor of
Li-like ions were carried out previously
\cite{heg75,ves80,lin93pra,yan01prl,yan02jpb,ind01hi}. However, in order to
reach the accuracy comparable to the one for H-like ions, a systematic
treatment of the QED corrections is required. In our previous publications
\cite{sha02pra,sha03nim}, we presented theoretical values for the $g$ factor
of Li-like ions including all presently known QED, nuclear, and
interelectronic-interaction corrections for a wide range of the nuclear-charge number
$Z$. The goal of the present investigation is to improve these theoretical
predictions by calculating several corrections that provide the leading
uncertainty to the theoretical values.

For low-$Z$ ions, the uncertainty of the $g$-factor values
\cite{sha02pra,sha03nim} was mainly determined by the error due to the
interelectronic-interaction correction. A part of this correction that is of
first order in $1/Z$ was evaluated in Ref.~\cite{sha02pra} within a rigorous
QED approach, whereas the remainder (of order $1/Z^2$ and higher) was
extracted from the variational calculations by Yan \cite{yan01prl,yan02jpb}
in case of low-$Z$ ions and from the many-body perturbation theory
calculations by Lindroth and Ynnerman \cite{lin93pra} for high-$Z$ ions. In
the present work, we evaluate the interelectronic-interaction correction of
order $1/Z^2$ and higher by a large-scale configuration-interaction
Dirac-Fock method.

For middle- and high-$Z$ ions, the main error of the theoretical values of
Ref.~\cite{sha02pra} originated from the binding QED correction. In that
work, it was evaluated to the leading order in $\az$ [namely, $(\az)^2$] and
to the zeroth order in $1/Z$. In the present investigation, we employ the
recent calculation of the self-energy term of the one-electron QED correction
to all orders in $\az$ \cite{yer04pra} and calculate the vacuum-polarization
term. We also evaluate the screened QED correction to its leading orders
in $\az$ and $1/Z$.

The relativistic units ($\hbar=c=1$) and the Heaviside charge unit
($\alpha=e^2/(4\pi), e<0$) are used in the paper.

\section{Formulation of the problem}
\label{basic}

The $g$ factor of an atom with a spinless nucleus is defined as
\be
\label{gfac1}
  g = \left . \frac{1}{\mub J}
  \frac{\partial E(\calH)}{\partial \calH} \right |_{\calH=0} =
      \left . \frac{1}{\mub J}
  \frac{\partial}{\partial \calH}
  \langle \Psi \mid H \mid \Psi \rangle \right |_{\calH=0},
\ee
where $H$ is the Hamiltonian of the atom in magnetic field $\bm{\calH}$,
$\calH = |\bm{\calH}|$ and the $z$ axis is directed along $\bm{\calH}$,
$E(\calH)$ is the energy of the state with the maximal ($M_J = J$) projection
of the angular momentum on the $z$ axis, and $\mub = |e|/(2m)$ is the Bohr
magneton. If the perturbation theory of first order in $\calH$ is used to
obtain the energy shift $\Delta E(\calH)$, Eq.~(\ref{gfac1}) can be
written as
\be
\label{gfac11}
  g = \frac{1}{\mub J}
  \frac{\Delta E(\calH)}{\calH}.
\ee
In what follows, we will use both definitions.

In the case of a high-$Z$ Li-like ion, in the zeroth-order approximation one
can neglect the interaction of the $2s$ electron with the closed $(1s)^2$
shell. We thus use the one-electron Dirac equation as the starting point in
our evaluation of the $g$ factor. The corresponding Hamiltonian is
\be
\label{hdirac}
  h = \balpha \cdot \bfp + \beta m + V + \vmagn\,,
\ee
where $V = - {\az}/{r}$ is the Coulomb potential induced by the nucleus and
$\vmagn = -e\balpha\cdot\bfA$ represents the interaction with the magnetic
field. For the homogeneous magnetic field the vector potential is $\bfA = 1/2
[\bm{\calH}\times\bfr]$ and thus $\vmagn = - (e/2)
\bm{\calH}\cdot[\bfr\times\balpha]$.

Evaluating the energy shift as the expectation value of $\vmagn$ with the
Dirac wave functions, we obtain the lowest-order $g$-factor value for the
$2s$ state,
\be
\label{gdirac}
  \gdirac = \frac23 \left(1+\sqrt{2+2\gamma}\right)
  = 2 - \frac{(\az)^2}{6} + \dots,
\ee
where $\gamma = \sqrt{1-(\az)^2}$. Various corrections to $\gdirac$ arise due
to the interelectronic interaction ($\dgint$),  one-electron QED effects
($\dgqed$), the screened QED effects ($\dgsqed$), and nuclear effects
($\Delta g_{\rm nuc}$). We thus write the total theoretical value for the $g$
factor as
\be
  g = \gdirac + \dgint + \dgqed + \dgsqed + \Delta g_{\rm nuc}\,,
\ee
with the corrections $\dgint$, $\dgqed$, $\dgsqed$, and $\Delta g_{\rm nuc}$
evaluated in Sections \ref{ieint}, \ref{qed}, \ref{sqed}, and \ref{sec:nucl},
respectively.

%

\section{Interelectronic interaction}
\label{ieint}

To evaluate the interelectronic-interaction effects, we start with
the relativistic Hamiltonian in the no-pair approximation, 
\begin{equation}
H = \Lambda_{+}\Bigl( \sum_j h(j) +\Hint\Bigr) \Lambda_{+}, \,\,\,\,\,\,\,\, \Hint = \HC
+ \HB\,, \label{ham1}
\end{equation}
where
$h$ is the one-particle Dirac Hamiltonian (\ref{hdirac}), the index
$j=1\ldots N$ numerates the electrons, $\HC$ is the Coulomb interaction,
$\HB$ is the Breit interaction, and $\Lambda_{+}$ is the projector on the
positive-energy states, which is the product of the one-electron projectors
$\lambda_{+}(i)$,
\begin{equation}
\Lambda_{+} = \lambda_{+}(1) \cdot \cdot \cdot \lambda_{+}(N)\,.
\label{projec1}
\end{equation}
Here,
\begin{equation}
\lambda_{+}(i) = \sum_{n} \mid u_n(i) \rangle \langle u_n(i) \mid\,,
\label{projec2}
\end{equation}
where $u_n(i)$ are the positive-energy eigenstates of an effective one-particle
Hamiltonian $h_u$. 
Two different Hamiltonians $h$ and $h^{\rm HFD}$ were considered as the
operator $h_u$, where $h$ is the one-electron Dirac Hamiltonian
(\ref{hdirac}) and $h^{\rm HFD}$ is the Hartree-Fock-Dirac operator without
the Breit interaction but including the interaction with the external field
\cite{Sucher,Mittleman}. 
However, since $h^{\rm HFD}$ corresponds to a better zero approximation
and, therefore, provides much faster convergence,
all the final results were obtained with this Hamiltonian.
In both cases the functions $u_n$ and hence the
projector $\Lambda_{+}$ depend on the external magnetic field. Therefore, the
derivative with respect to $\calH$ in Eq.~(\ref{gfac1}) contains not only the
derivative of the one-electron part of the Hamiltonian (\ref{ham1}), but also
the derivative of the projector $\Lambda_{+}$. The derivative of
$\Lambda_{+}$ can be obtained explicitly and one can show that its
contribution to the $g$ factor is equivalent to the contribution of the
negative-energy states.

In our computational approach, we obtain the $g$-factor value by using
Eq.~(\ref{gfac1}) in the finite-difference approximation,
\begin{equation}
  g = \frac{1}{\mub J}
  \frac{E(\delta\calH)-E(-\delta\calH)}{2\,\delta\calH} +
  O[(\delta\calH)^3]\,,
\label{gfac2}
\end{equation}
since the Hellman-Feynman theorem is not exactly fulfilled for the
approximate wave functions. The optimal choice of the finite difference
$\delta\calH$ was found to be $0.001\, ({\rm a.u.}/\mub)$. We checked that
scaling this value by a factor of $2$ does not influence our results for the
$g$ factor.

In order to determine the space of one-electron functions $\left\{
\varphi_j\right\}_{j = 1}^M$, we employed the combined Dirac-Fock (DF)
($j=1,\dots, m$) and the Dirac-Fock-Sturm (DFS) ($j=m+1,\dots, M$) basis set.
The index $j$ here enumerates different occupied and vacant one-electron
states. The external magnetic field and the Breit interaction were not
included in the DF and DFS operators, when the basis set was generated. For
the occupied atomic shells, the orbitals $\varphi_{j}$ were obtained by the
restricted DF method, based on a numerical solution of the radial DF
equations. The vacant orbitals ${\varphi}_{j}$ ($j=m+1,\dots, M$) were
obtained by solving the Dirac-Fock-Sturm equations
\begin{equation}
\left [ {h}^{\rm DF} - \varepsilon_{j_0}
\right] \tilde \varphi_{j}
 = \lambda_{j} W(r) \tilde \varphi_{j}\,, \qquad j=m+1,\dots, M,
\label{Sturm1}
\end{equation}
where $ {h}^{\rm DF}$ is the Dirac-Fock operator, $\varepsilon_{j_0}$ is the
one-electron energy of the occupied DF orbital  $\varphi_{j_0}$, and  $W(r)$
is a constant-sign weight function. The parameter $\lambda_{j}$ in
Eq.~(\ref{Sturm1}) can be considered as an eigenvalue of the Sturmian
operator. If  $W(r) \to 0$ at $r \to \infty$, all Sturmian functions
$\varphi_{j}$ have the same asymptotics at $r \to \infty$. It is clear that
for $\lambda_j=0$ the Sturmian function coincides with the reference DF
orbital $\varphi_{j_0}$. The widely used choice of the weight function is
$W(r)=1/r$, which leads to the well-known ``charge quantization". In the
relativistic case this choice is not very successful. In our calculations we
used the following weight function
\begin{equation}
W(r) = \frac{1 - \exp[-(\alpha r)^2]}{(\alpha r)^2}\,, \label{sturm2}
\end{equation}
which, unlike $1/r$, is regular at the origin. It is well known that the
Sturmian operator is Hermitian and, contrary to the Fock operator, does not
have the continuum spectra. Therefore, the set of the Sturmian eigenfunctions
including the negative-energy states forms the discrete and complete basis
set in the space of one-electron wave functions. This basis set is orthogonal
with the weight function $W(r)$.

To generate the one-electron wave functions $\psi_n$, we used the
unrestricted DF method in the joined DF and DFS basis,
\begin{equation}
\psi_n  = \sum_{j} C_{jn} \varphi_j\,. \label{hf1}
\end{equation}
The coefficients $C_{jn}$ were obtained by solving HFD matrix equations
\begin{equation}
{\hat F} {\bf C}_n  = \varepsilon_n {\hat S} {\bf C}_n,
\label{hf2}
\end{equation}
where ${\hat F}$ is the Dirac-Fock matrix in the joined basis of DF and DFS
orbitals of a free ion. The external magnetic field was included in the
${\hat F}$ matrix, whereas the Breit interaction was not. The matrix ${\hat
S}$ in Eq.~(\ref{hf2}) is nonorthogonal, since the DFS orbitals are not
orthogonal in the usual sense. The negative-energy DFS functions were
included in the total basis set. Eq.~(\ref{hf2}) was used to generate the
whole set of orthogonal one-electron wave functions $\psi_n (n=1,\dots M)$,
including all vacant states.

It should be noted that even if the external magnetic field $\calH$ is equal
to zero, the set of one-electron functions $\psi_n$ differs from the set of
basis functions $\varphi_j$. For the occupied states, the unrestricted DF
method accounts for the core-polarization effects (the spin polarization in
our case), in contrast to the restricted DF method. For the vacant states the
difference is more significant, since the DF and DFS operators are
essentially different.

The large-scale configuration-interaction  Dirac-Fock (CI-DF) method was used
to solve the Dirac-Coulomb-Breit equation in the external magnetic field
\begin{equation}
H \Psi(\gamma M_J) = E(\gamma M_J)\, \Psi(\gamma M_J)
\end{equation}
where $H$ is the non-pair Hamiltonian (\ref{ham1}). The many-electron wave
function $\Psi(\gamma M_J)$ with quantum numbers $\gamma$ and $M_J$ was
expanded in terms of a large number of the Slater determinants (SD) with the
same projection $M_J$ of the total angular momentum $J$
\begin{equation}
\Psi(\gamma M_J) = \sum_{\alpha} c_{\alpha}(\gamma M_J)\, {\rm
det}_{\alpha}(M_J). \label{expan1}
\end{equation}
The configuration state functions (CSFs) with angular momentum $J$ were not
used in our calculations, since the Hamiltonian $H$ contains the interaction
with the external magnetic field and, therefore, does not commute with the
operator $J^2$. The Slater determinants are constructed from the one-electron
wave functions $\psi_n$ (\ref{hf1}). The same orbitals were used in
Eq.~(\ref{projec2}) in order to construct the projector $\Lambda_{+}$. The
basis of one-electron functions used in our calculations was
$12s~11p~10d~6f~4g~2h~1i$. The set of the SD in expansion (\ref{expan1}) was
generated including all single, double, and triple excitations.
The total number of SD was 552359.
The results of the calculation are presented in Table~\ref{tab:int}.
The interelectronic-interaction correction $\dgdf$ is the difference of the
result obtained by Eq.~(\ref{gfac2}) for the point nuclear model and the
Dirac $g$-factor value $\gdirac$ (\ref{gdirac}).

The result for the $\dgdf$ correction obtained by the CI-DF method can be
improved by employing a rigorous QED treatment of the part of this correction
that is of order $1/Z$, which was presented in Ref.~\cite{sha02pra}. In order
to combine two different treatments, we isolate the contribution of order
$1/Z^2$ and higher from the $\dgdf$ correction by subtracting the value of
the $1/Z$ term calculated in the Breit approximation. 
The resulting ``higher-order" correction ($\dgho$) is
listed in the third column of Table~\ref{tab:int}. The numerical results for
the interelectronic-interaction correction of first order in $1/Z$
($\dgexact$) are taken from Ref.~\cite{sha02pra} and listed in the fourth
column. This contribution was evaluated in framework of QED and utilizing
the Fermi model for the nuclear-charge distribution. Total results $\dgint =
\dgexact+\dgho$ are presented in the last column. The error bars indicated
represent a quadratical sum of a numerical error and an estimation of omitted
terms, i.e., contributions beyond the Breit approximation to $\dgho$. They
were estimated as $(\az)^2\, \dgho$.

To compare our results with the corresponding Yan's calculations
\cite{yan01prl,yan02jpb}, which account for the lowest-order $(\sim 
\alpha^2)$ relativistic effects, we have isolated the $\alpha^2$
contribution in our CI-DF calculation. It was done by four times increase
of the velocity of light (in atomic units) and by an extrapolation
of the obtained results  (with the $\alpha^2$ factor isolated) 
 to the $c =  \infty$ limit.
  Table~\ref{tab:intnr},
which presents the related comparison for $g-2$ values,
 shows that for $Z>5$ the contribution
of the higher-order relativistic  effects is much larger than the
 difference between our $\alpha^2$ results and those of Yan.

%
\section{One-electron QED corrections}
\label{qed}

To zeroth order in $1/Z$, the QED correction for the ground state of a
Li-like ion is given by the one-electron QED contribution evaluated for the
$2s$ Dirac state. This correction is represented by a perturbation expansion
in the fine-structure constant $\alpha$,
\begin{equation}
\dgqed = \dgqed^{(1)}+ \dgqed^{(2)}+ \ldots\,,
\end{equation}
where the superscript indicates the order in $\alpha$.

The first-order QED correction is given by the sum of the self-energy and
vacuum-polarization contributions, $\dgqed^{(1)} = \dgse^{(1)} +
\dgvp^{(1)}$. The self-energy correction for the $2s$ state was recently
calculated to all orders in $\az$ in Ref.~\cite{yer04pra}. The corresponding
results are listed in the second column of Table~\ref{tab:qed}. The
vacuum-polarization correction consists of two parts that can be thought to
originate from the first-order vacuum-polarization diagram with the magnetic
interaction inserted into the external electron line (the {\it electric-loop}
contribution $\dgvpe$), and into the vacuum-polarization loop (the {\it
magnetic-loop} contribution $\dgvpm$). The electric-loop contribution is
calculated in the present work by utilizing the known expression for the
Uehling potential and approximate formulas for the Wichmann-Kroll potential
taken from Ref.~\cite{fai90jpb}. The results of the calculation are presented
in the third and fourth columns of Table~\ref{tab:qed}. The remaining magnetic-loop
correction is known to vanish in the Uehling approximation, and its
contribution is small as compared to the electric-loop part. This correction
was calculated to all orders in $\az$ only for the $1s$ state
\cite{per97pra,bei00pra}. Because of this, we employ the analytical result of
Ref.~\cite{kar02plb} for its leading contribution in $\az$, which reads (for
an $ns$ state),
\be \label{eq:vpmagn} \dgvpm = \frac{7}{216}\,
\frac{\alpha(\az)^5}{n^3}\,.
\ee
The corresponding results are listed in the fifth column of
Table~\ref{tab:qed}. The uncertainty of the magnetic-loop contribution is
estimated by comparison of the all-order numerical results of
Ref.~\cite{bei00pra} for the $1s$ state with the lowest-order analytical
result (\ref{eq:vpmagn}).

QED corrections of higher orders in $\alpha$ have not been calculated to all
orders in $\az$ up to now. Two first terms of their $\az$ expansion can be
represented by the free-electron $g-2$ factor multiplied by a relativistic
kinematical factor \cite{gro70,cza01pra,kar:01:hydr,sha02pra}. The result
yields (for an $ns$ state)
\begin{equation} \label{eq:qedho1}
  \dgqed^{(i)} = 2\,\left(\frac{\alpha}{\pi} \right)^i A^{(i)}\,
        \left(1 + \frac{(\az)^2}{6\,n^2} \right)\,,
\end{equation}
where $2(\alpha/\pi)^i A^{(i)}$ is the contribution of order
$\alpha^i$ to the free-electron $g$ factor. Its numerical values (see
\cite{hughes99rmp,kinoshita:03:prl} and references therein) are
\begin{eqnarray}
 A^{(1)} &=& \frac12\,, \\
 A^{(2)} &=& -0.328\,478\,965\ldots\,, \\
 A^{(3)} &=& 1.181\,241\,456\ldots\,, \\
 A^{(4)} &=& -1.7366(384)\,.
\end{eqnarray}

Recently, a part of the two-loop QED contribution of order $\alpha^2 (\az)^4$
was evaluated in Ref.~\cite{pachucki:04} with the result
\begin{equation} \label{eq:qedho2}
  \dgqed^{(2)}({\rm h.o.}) = \left(\frac{\alpha}{\pi} \right)^2
 \frac{(\az)^4}{n^3} \left\{ \frac{56}{9}\ln[(\az)^{-2}] + a_{40} \right\} \,,
\end{equation}
where the numerical values of the coefficient term are $a_{40}^{(2)}(1s) =
-18.477\,948\,664\,(1)$ and $a_{40}^{(2)}(2s) = -19.781\,820\,939\,(1)$. This
expression accounts for the complete logarithmic dependence in this order and
the dominant part of the constant term. We observe, however, that in the
one-loop case, inclusion of the term of the order $(\az)^4$ is meaningful for
sufficiently small values of $Z$ only. In the high-$Z$ region, addition of
this term makes the $\az$-expansion results deviate more from the ``exact"
numerical values. We thus include the contribution (\ref{eq:qedho2}) for
$Z\le30$ only. The relative uncertainty of the two-loop binding QED
correction for $Z\le30$ was estimated as the ratio of the part of the
one-loop QED correction that is of order $(\az)^5$ and higher to the part
that is within the $(\az)^4$ approximation \cite{pachucki:04}, multiplied by
a factor of 4. For $Z>30$, we estimate the relative uncertainty by the ratio
of the part of the one-loop QED correction that is of order $(\az)^4$ and
higher to the part that is within the $(\az)^2$ approximation, multiplied by
a factor of 2.

%
\section{Screened QED corrections}
\label{sqed}

In this section we investigate the influence of the interelectronic
interaction on the QED effects, known also as the ``screening" of QED
corrections. 
To derive the screened QED correction of first order in $1/Z$, we apply an
approximate method, which yields the complete result to the order
$(\az)^2$. 
Following Hegstrom \cite{heg73}, we adopt the Hamiltonian
\be
\label{ham2}
  H &=& \sum_j h(j)+  \sum_j h^{\rm rad}(j)
+ \Hint,
\ee
where $h$ is defined by Eq. (\ref{hdirac}),
$\Hint$ incorporates the interelectronic interaction within the Breit
approximation (\ref{ham1}), and $h^{\rm rad}$  accounts for the interaction
of the anomalous magnetic moment of electron with the magnetic
and electric 
fields,
\be
\label{addpauli}
  \hrad &=&
  \frac{\gfree-2}{2}\, \mub \, \left[\beta (\bSigma\cdot{\bm{\calH}})-
  i\beta(\balpha\cdot{\bm{\calE}}) \right].
\ee
Here
\be
\gfree = 2\, \sum_i \left(\frac{\alpha}{\pi}\right)^i A^{(i)}
\ee
is the free-electron $g$ factor and $\bSigma =
\Dmatrix{\bsigma}{0}{0}{\bsigma}$.
The magnetic field $\bm{\calH}(j)$ acting on the
$j$th electron in Eq.~(\ref{addpauli}) includes the external homogeneous
magnetic field $\bm{\calH}$ and the field induced by the other electrons
\be
  \bm{\calH}(j) = \bm{\calH}
  + \sum_{k \neq j}
  \frac{e}{4\pi}\frac{\balpha_k\times\bfr_{jk}}{r_{jk}^3}\,,
\ee
where $\bfr_{jk} = \bfr_j - \bfr_k$. The electric field $\bm{\calE}(j)$
includes the fields induced by nucleus and by the other electrons
\be
  \bm{\calE}(j) = \frac{|e|Z}{4\pi} \frac{\bfr_j}{r_j^3}
  + \sum_{k \neq j} \frac{e}{4\pi} \frac{\bfr_{jk}}{r_{jk}^3}\,.
\ee
We divide the contribution arising from $\hrad$ into three parts,
\be
\Hrada &=& \frac{\gfree-2}{2} \, \mub \, \sum_j \beta_j (\bm{\calH} \cdot
\bSigma_j)\,,
\\
\Hradb &=& \frac{\gfree-2}{2} \, \mub \, \frac{|e|Z}{4\pi} (-i)
         \sum_j \beta_j \frac{\balpha_j \cdot \bfr_j}{r_j^3}\,,
\\
\Hradc &=& \frac{\gfree-2}{2} \, \mub \, \frac{e}{4\pi} \sum_{j \neq k}
    \left ( \beta_j \bSigma_j \cdot \frac{\balpha_k \times \bfr_{jk}}{r_{jk}^3}
            - i \beta_j \frac{\balpha_j \cdot \bfr_{jk}}{r_{jk}^3} \right )
\,.
\ee
The matrices $\balpha_j$, $\beta_j$, $\bSigma_j$ here act on the spinor
variables of the $j$th electron. To first order in $1/Z$, the screened QED
correction to the $g$ factor can be now derived by the standard
Rayleigh-Schr\"odinger perturbation theory, separating corrections linear in
the magnetic field  $\calH$, which are of first order in the parameter $1/Z$.
The contributions of interest can be conventionally represented by the
following combinations:
\be
\label{de1}
  \Delta E_1 &\sim& \Hrada \times \Hint + {\rm(permutations)}\,,
\ee
\be
\label{de2}
  \Delta E_2 &\sim& \Hradb \times \vmagn \times \Hint + {\rm(permutations)}\,,
\ee
\be
\label{de3}
  \Delta E_3 &\sim& \Hradc \times \vmagn + {\rm(permutations)}\,.
\ee
After angular integration, the summation over the complete Dirac-Coulomb
spectrum was performed by the finite basis set method with basis functions
constructed from $B$ splines \cite{joh86,joh88}. The numerical results for
the screened QED correction of first order in $1/Z$, $\dgsqedz$, are presented
in Table~\ref{tab:sqed} in terms of the function $R(\az)$, defined as
\be
\label{dgsqed}
  \dgsqedz = (\gfree-2)\, \frac{(\az)^2}{Z}\, R(\az)\,.
\ee
The terms $R_{i}(\az)$ with $i=1,2$, and 3 are induced by
Eqs.~(\ref{de1})-(\ref{de3}), respectively.

The screened QED correction of higher orders in $1/Z$ should be accounted for
when considering low-$Z$ ions. We extract this correction from the recent
evaluations by Yan \cite{yan01prl,yan02jpb}, which were performed on
nonrelativistic wave functions but with the interelectronic interaction taken
into account to all orders in $1/Z$. 
Yan's results for the screened QED correction can
be represented in the form
\be \label{eq:yan}
  \dgsqed({\rm Yan}) = (\gfree-2)\, (\az)^2\,
  \left [ \frac{1}{Z}R(0) + \frac{1}{Z^2}Q(0) + \dots \right ]\,,
\ee
where $R(0) = -274/2187$. The functions $R, Q, \dots$ here do not have any
dependence on $Z$ since Yan's calculations are based on the nonrelativistic
form of the Hamiltonian (\ref{ham2}). We obtain the numerical value of $Q(0)$
by fitting Yan's results for $Z=3$--$12$ to the form (\ref{eq:yan}), which
yields $Q(0) = 0.071(1)$. With this value of $Q(0)$, formula (\ref{eq:yan})
 was used to estimate the
higher-order screened QED correction for $Z > 12$.

In Table~\ref{tab:sqedtot} we present the results for the screened QED
correction of first order in $1/Z$ ($\dgsqedz$) and of higher orders in $1/Z$
($\dgsqedzz$). The error of the term $\dgsqedz$ was estimated as
 the part of the one-electron QED correction that is beyond the
$(\az)^2$ approximation, multiplied by a factor $3/Z$.
The uncertainty ascribed to the $\dgsqedzz$  contribution was evaluated
as $3\,(\az)^2\,\dgsqedzz\,$.

%
\section{Nuclear effects}
\label{sec:nucl}

In this section we briefly summarize the known results for the nuclear
effects on the $g$ factor of Li-like ions. The correction to the Dirac
$g$-factor value due to the extended nuclear size is relatively simple. For
high-$Z$ ions, it is evaluated numerically by employing the Fermi model for
the nuclear-charge distribution. The uncertainty is estimated by taking the
difference of the results obtained for the Fermi and sphere nuclear models.
For low-$Z$ ions, to a good accuracy, this correction can be evaluated by
 a simple analytical
formula obtained in Ref.~\cite{gla02pla},
\be \label{eq:fns}
  \dgns = \frac{1}{3} (\az)^4 m^2 \langle r^2 \rangle
  \left [ 1 + (\az)^2 \left ( \frac{35}{16} - C -
  \frac{\langle r^2 \ln(\az mr) \rangle} {\langle r^2 \rangle} \right ) \right] \,.
\ee
Here $C = 0.5772156649 \dots$ is the Euler constant and the expectation value
has to be evaluated with the proper nuclear-charge density.

Systematic QED theory for the nuclear-recoil effect on the atomic $g$ factor
to the first order in $m/M$ and to all orders in $\az$ was developed in
Ref.~\cite{sha02pra:rec}. The one-electron
recoil correction derived in that work is expressed
as the sum of the lower-order and the higher-order term. The first one can be
calculated analytically to yield for the $2s$ state (cf. \cite{sha02pra:rec})
\be
\label{dgrec}
  \Delta g_{\rm rec,L} = \frac{m}{M}\frac{(\az)^2}{4}
  \left\{ 1 + (\az)^2 \frac{14+6\gamma+12\sqrt{2(1+\gamma)}}
  {3(1+\gamma)^2 \bigl[ 2+\sqrt{2(1+\gamma)}\, \bigr] ^2} \right\} \,,
\ee
where $\gamma = \sqrt{1-(\az)^2}$. The higher-order term $\Delta g_{\rm
rec,H}$ 
 was calculated
numerically in Ref.~\cite{sha02} for the $1s$ state only. 
These numerical results showed the   $(\az)^5$
behaviour of this term at low $Z$.
We estimate the
relative uncertainty of the result (\ref{dgrec}) due to neglecting the
higher-order term as the ratio $\Delta g_{\rm rec,H}/\Delta g_{\rm rec,L}$
for the $1s$ state, multiplied by a factor of $1.5$.

The contribution of the two-electron recoil effect can be extracted from the
results of Yan \cite{yan01prl,yan02jpb}. For $Z=3-12$, it was evaluated as the
difference of Yan's result of order $\alpha^2 m/M$ and the nonrelativistic
limit of Eq.~(\ref{dgrec}). Fitting this difference to the form
\be \label{eq:recoil:yan}
  \dgrec^{\rm two-el} =\frac{m}{M}\frac{(\az)^2}{4}
  \left [ \frac{1}{Z}C + \frac{1}{Z^2}D \dots \right ]\,
\ee
yields $C = -3.3(2) $. With this value of $C$, formula (\ref{eq:recoil:yan})
 was used to estimate the two-electron recoil
correction  for  $Z>12$.

Finally, we note that the nuclear polarization effect on the atomic 
$g$ factor was evaluated in Ref. \cite{nef02prl}.

%

\section{Results and discussion}

In Table~\ref{tab:total}, we present the individual contributions to the $g$
factor of the ground state of Li-like ions. The Dirac point-nucleus value is
obtained by Eq.~(\ref{gdirac}). The finite nuclear size correction to this
value is evaluated by Eq.~(\ref{eq:fns}) for low-$Z$ ions and by a direct
numerical solution of the Dirac equation for high-$Z$ ions. The
interelectronic-interaction correction $\dgint$ is the sum of the part of
first order in $1/Z$, $\dgexact$, obtained in framework of QED and of the
higher-order part, $\dgho$, evaluated by the CI-DF method, as described in
Section~\ref{ieint}. The QED correction of order $\alpha$ is the sum of the
one-electron self-energy and vacuum-polarization terms presented in
Table~\ref{tab:qed}. The QED correction of order $\alpha^2$ and higher
incorporates the known terms of the $\az$ expansion, as
explained in Section~\ref{qed}. The screened QED correction is discussed in
Section~\ref{sqed}, the corresponding results are taken from
Table~\ref{tab:sqedtot}. The nuclear recoil correction is obtained as
explained in Section ~\ref{sec:nucl}. For lead and uranium, we include also the
nuclear-polarization correction calculated in Ref.~\cite{nef02prl}.

Table~\ref{tab:total} demonstrates a significant improvement achieved
comparing to our previous evaluations \cite{sha02pra,sha03nim}. The
uncertainty of the presented theoretical values for carbon and oxygen is 2
and 3 times better than those of Ref.~\cite{sha03nim}, respectively; whereas
for uranium the accuracy is improved by two orders of magnitude. Progress in
the high-$Z$ region is mainly due to the evaluation of the one-loop QED
corrections to all order in $\az$, while for low-$Z$ ions it is largely due
to the interelectronic-interaction correction and the screened QED
correction.

The accuracy of the theoretical values in Table~\ref{tab:total} is several
parts in $10^{-8}$ for low-$Z$ ions and a few parts in $10^{-7}$ for
middle-$Z$ ions. It decreases further with $Z$ increasing and reaches $5
\times 10^{-6}$ for uranium. So, despite the achieved improvement, the
accuracy for Li-like ions is still significantly lower than that for H-like
ions \cite{yer02prl,sha02} and also than the precision that can be
presently addressed in experiments \cite{her00,haf00,ver04}. In particular,
the nearest aim of experimental investigations of the Mainz-GSI collaboration
is the $g$ factor of Li-like calcium. The anticipated
experimental accuracy is at the ppb level, which can be compared with the
relative theoretical error of $ 10^{-7}$ from Table~\ref{tab:total}.

The uncertainty of the present theoretical values is mainly defined by the
interelectronic-interaction correction and by the screened QED correction.
 An improvement in the theoretical description 
of the interelectronic-interaction effects
can be
achieved by a rigorous QED treatment of the part of order $1/Z^2$ and by
calculating the remainder within the Breit approximation. Such a program has
been carried out for the Lamb shift in Li-like ions
\cite{zhe00,yerokhin:00:prl,sha01:hi,andreev:01}.
 However, a similar calculation for the $g$
factor is going to be significantly more difficult due to the presence of the
external magnetic interaction and requires a further development of methods
of calculational QED. 
As to the screened QED correction, in the present work it was
calculated to its leading order in $\az$ only. 
As a first step beyond this approximation, which can improve 
the results for this correction in the high-$Z$ region,
one may consider an evaluation of
the one-loop QED corrections in an effective potential that partly accounts
for the interelectronic-interaction effects \cite{ind90,sap02}.
These two topics will be the
subjects of our subsequent investigations.

In summary, we have presented calculations of the
interelectronic-interaction, one-electron QED, and screened QED corrections
to the $g$ factor of the ground state of Li-like ions. This resulted in a
significant improvement of theoretical predictions in a wide range of the
nuclear-charge values $Z$. We have analyzed also perspectives for further
progress in the theoretical description of these systems and for probing the
QED effects in future experiments.

\acknowledgments

Valuable conversations with A.N. Artemyev, T. Beier, S. Djekic, P. Indelicato,
H.-J. Kluge, W. Quint, and G. Werth are gratefully
acknowledged. This work was supported in part by RFBR (Grants No.
04-02-17574, 03-02-33253a), by the Russian Ministry of Education (Grant No.
E02-3.1-49), and by INTAS-GSI (Grant No. 03-54-3604).
D.A.G. thanks for support from the foundation ``Dynasty'', the Russian
Ministry
of Education (Grant No. A03-2.9-261), the Max-Planck-Institut f\"ur Physik
Komplexer Systeme and from DAAD during his research stays at TU Dresden.
 The work of V.M.S. was supported by the Alexander
von Humboldt Stiftung. V.A.Y. acknowledges the support of the foundation
``Dynasty". G.P. and G.S. acknowledge financial support by the BMBF, DFG, and
GSI.


\begin{table}
\caption{The interelectronic-interaction correction $\dgint =
\dgexact+\dgho$. The contributions $\dgdf$ and $\dgho$ are calculated for the
point nucleus, whereas the correction $\dgexact$ -- for the Fermi model of
the nuclear-charge distribution. All values shown in units of $10^{-6}$.
\label{tab:int}} \vspace{0.5cm}
\begin{tabular}{rr@{}lr@{}lr@{}lr@{}l}
\hline
\hline
$Z$ &
\multicolumn{2}{c}{$\dgdf$} &
\multicolumn{2}{c}{$\dgho$} &
\multicolumn{2}{c}{$\dgexact$} &
\multicolumn{2}{c}{$\dgint$}\\
\hline
3  &   61.&591&    $-$7.&082&   68.&676&   61.&594 (10) \\
4  &   84.&822&    $-$6.&752&   91.&591&   84.&839 (12) \\
5  &  107.&806&    $-$6.&679&  114.&493&  107.&814 (14) \\
6  &  130.&745&    $-$6.&661&  137.&419&  130.&758 (19) \\
8  &  176.&627&    $-$6.&662&  183.&320&  176.&658 (30) \\
10 &  222.&571&    $-$6.&673&  229.&301&  222.&628 (42) \\
12 &  268.&604&    $-$6.&682&  275.&385&  268.&703 (55) \\
14 &  314.&744&    $-$6.&689&  321.&592&  314.&903 (74) \\
16 &  361.&011&    $-$6.&695&  367.&939&  361.&244 (94) \\
18 &  407.&422&    $-$6.&699&  414.&451&  407.&75 (12) \\
20 &  453.&994&    $-$6.&702&  461.&148&  454.&45 (14) \\
24 &  547.&70&    $-$6.&71&  555.&18&  548.&48 (21) \\
32 &  737.&93&    $-$6.&71&  746.&46&  739.&75 (37) \\
54 & 1291.&39&    $-$6.&81& 1306.&22& 1299.&4 (1.1) \\
82 & 2120.&76&    $-$7.&64& 2148.&29& 2140.&7 (2.7) \\
92 & 2481.&67&    $-$8.&48& 2509.&84& 2501.&4 (3.8) \\
\hline
\end{tabular}
\end{table}

\begin{table}
\caption{The comparison of the present CI-DF calculations of the $g$ factor
and the values obtained by Yan \cite{yan01prl,yan02jpb}.
The second column presents our $g-2$ values, incorporating
the effects of binding and electron correlation.
The $\alpha^2$ limit is shown in the third column.
Yan's contribution to the $g-2$ value of order $\alpha^2$ is listed in third column.
All numbers are in units of $10^{-6}$
\label{tab:intnr}} \vspace{0.5cm}
\begin{tabular}{rr@{}lr@{}lr@{}l}
\hline
\hline
$Z$ &
\multicolumn{2}{c}{\,\,$g-2\;$ (CI-DF)\,\,} &
\multicolumn{2}{c}{\,\,$g-2\;$ (CI-DF, $\sim \alpha^2$ )\,\,} &
\multicolumn{2}{c}{\,\,$g-2\;$ (Yan \cite{yan01prl,yan02jpb})\,\,}\\
\hline
3   &    -18.&298  &    -18.&293  &    -18.&283  \\
4   &    -57.&220  &    -57.&206  &    -57.&209  \\
5   &   -114.&167  &    -114.&123     &   -114.&132  \\
6   &   -188.&955  &   -188.&845  &   -188.&859  \\
8   &   -391.&994  &   -391.&580  &   -391.&599  \\
10  &   -666.&382  &   -665.&326  &   -665.&348  \\
12  &  -1012.&504  &  -1010.&072  &  -1010.&096  \\
\hline
\end{tabular}
\end{table}


\begin{table}
\caption{One-electron self-energy and vacuum-polarization corrections of
first order in $\alpha$. $\dgvpeu$ and $\dgvpewk$ are the Uehling and the
Wichmann-Kroll part of the electric-loop vacuum-polarization contribution,
respectively; $\dgvp^{(1)} = \dgvpeu+ \dgvpewk+\dgvpm$. All contributions are
calculated with the point nuclear model. Units are $10^{-6}$.
\label{tab:qed}} \vspace{0.5cm}
\begin{tabular}{rr@{}lr@{}lr@{}lr@{}lr@{}lr@{}l}
\hline
\hline
$Z$ & \multicolumn{2}{c}{$\dgse^{(1)}$} & \multicolumn{2}{c}{$\dgvpeu$} &
\multicolumn{2}{c}{$\dgvpewk$} & \multicolumn{2}{c}{$\dgvpm$} &
\multicolumn{2}{c}{$\dgvp^{(1)}$} & \multicolumn{2}{c}{$\dgqed^{(1)}$}\\
\hline

4 &

2322.&905 1 (4) &
$-$0.&000 215 &
   0.&000 000 &
   0.&000 001 &
$-$0.&000 214 &
2322.&904 9 (4) \\

6 &

2323.&018 3 (6) &
$-$0.&001 070 &
   0.&000 000 &
   0.&000 005 &
$-$0.&001 065 &
2323.&017 2 (6) \\

8 &

2323.&185 (1) &
$-$0.&003 326 &
   0.&000 002 &
   0.&000 020 &
$-$0.&003 304 &
2323.&182 (1) \\

10 &

2323.&413 (2) &
$-$0.&008 00  &
   0.&000 01  &
   0.&000 06  &
$-$0.&007 93  &
2323.&405 (2) \\

12 &

2323.&707 (2) &
$-$0.&016 35  &
   0.&000 02  &
   0.&000 15 (1) &
$-$0.&016 18 (1) &
2323.&691 (2) \\

14 &

2324.&074 (3) &
$-$0.&029 92 &
   0.&000 06 &
   0.&000 33 (2) &
$-$0.&029 53 (2) &
2324.&044 (3) \\

16 &

2324.&520 (3) &
$-$0.&050 48  &
   0.&000 12  &
   0.&000 64 (4) &
$-$0.&049 71 (4) &
2324.&470 (3) \\

18 &

2325.&052 (5) &
$-$0.&080 06  &
   0.&000 25  &
   0.&001 16 (8) &
$-$0.&078 66 (8) &
2324.&973 (5) \\

20 &

2325.&674 (5) &
$-$0.&121 0   &
   0.&000 5   &
   0.&002 0 (2) &
$-$0.&118 6 (2) &
2325.&555 (5) \\

24 &

2327.&225 (5) &
$-$0.&247 5   &
   0.&001 3   &
   0.&004 9 (5) &
$-$0.&241 3 (5) &
2326.&984 (5) \\

32 &

2331.&726 (6) &
$-$0.&772     &
   0.&007     &
   0.&021 (3) &
$-$0.&745 (3) &
2330.&981 (7) \\

54 &

2358.&184 (9) &
$-$6.&652     &
   0.&138     &
   0.&28 (8)  &
$-$6.&23 (8)  &
2351.&95 (8) \\

82 &

 2456.&245 (9) &
$-$48.&266     &
    1.&886     &
    2.&3 (8)   &
$-$44.&1 (8)   &
 2412.&1 (8)  \\

92 &

 2532.&207 (9) &
$-$93.&309     &
    4.&260      &
    4.&0 (1.5)  &
$-$85.&0 (1.5)  &
 2447.&2 (1.5) \\
\hline
\hline
\end{tabular}
\end{table}


\begin{table}
\caption{Individual contributions to the screened QED correction of order
$1/Z$ in terms of the function $R(\az)$ defined by Eq.~(\ref{dgsqed}). The
terms $R_{i}$ with $i=1,2,3$ are induced by Eqs.~(\ref{de1})-(\ref{de3}),
respectively. The results in the last column are obtained for the
extended-charge nucleus, whereas all the other results correspond to the
point nuclear model. \label{tab:sqed}} \vspace{0.5cm}
\begin{tabular}{rr@{}lr@{}lr@{}lr@{}lr@{}l}
\hline
\hline
$Z$ &
\multicolumn{2}{c}{$R_{1}$} &
\multicolumn{2}{c}{$R_{2}$} &
\multicolumn{2}{c}{$R_{3}$} &
\multicolumn{2}{c}{$R(\az)$}       &
\multicolumn{2}{c}{$R_{\rm f.n.}(\az)$}\\
\hline
4 &
  0.&1134 &
 $-$0.&1139 &
 $-$0.&1253 &
 $-$0.&1258 &
 $-$0.&1258 \\
6 &
  0.&1135 &
 $-$0.&1144 &
 $-$0.&1254 &
 $-$0.&1264 &
 $-$0.&1264 \\
8 &
  0.&1135 &
 $-$0.&1153 &
 $-$0.&1254 &
 $-$0.&1272 &
 $-$0.&1272 \\
10 &
  0.&1136 &
 $-$0.&1163 &
 $-$0.&1255 &
 $-$0.&1283 &
 $-$0.&1283 \\
12 &
  0.&1136 &
 $-$0.&1176 &
 $-$0.&1256 &
 $-$0.&1296 &
 $-$0.&1296 \\
14 &
  0.&1137 &
 $-$0.&1192 &
 $-$0.&1258 &
 $-$0.&1312 &
 $-$0.&1312 \\
16 &
  0.&1138 &
 $-$0.&1210 &
 $-$0.&1259 &
 $-$0.&1331 &
 $-$0.&1331 \\
18 &
  0.&1140 &
 $-$0.&1231 &
 $-$0.&1261 &
 $-$0.&1352 &
 $-$0.&1352 \\
20 &
  0.&1141 &
 $-$0.&1255 &
 $-$0.&1263 &
 $-$0.&1377 &
 $-$0.&1377 \\
24 &
  0.&1144 &
 $-$0.&1311 &
 $-$0.&1268 &
 $-$0.&1435 &
 $-$0.&1435 \\
32 &
  0.&1152 &
 $-$0.&1465 &
 $-$0.&1280 &
 $-$0.&1593 &
 $-$0.&1590 \\
54 &
  0.&1188 &
 $-$0.&2324 &
 $-$0.&1336 &
 $-$0.&2473 &
 $-$0.&2436 \\
82 &
  0.&1274 &
 $-$0.&6196 &
 $-$0.&1491 &
 $-$0.&6412 &
 $-$0.&5568 \\
92 &
  0.&1324 &
 $-$1.&0501 &
 $-$0.&1588 &
 $-$1.&0765 &
 $-$0.&8103 \\
\hline
\hline
\end{tabular}
\end{table}

\begin{table}
\caption{The total screened QED correction. $\dgsqedz$ and $\dgsqedzz$ are
the screened QED contribution of first and of higher orders in $1/Z$,
respectively. All numbers are in units of $10^{-6}$. \label{tab:sqedtot}}
\vspace{0.5cm}
\begin{tabular}{rr@{}lr@{}lr@{}l}
\hline
\hline
$Z$ &
\multicolumn{2}{c}{$\dgsqedz$} &
\multicolumn{2}{c}{$\dgsqedzz$} &
\multicolumn{2}{c}{$\dgsqed$}\\
\hline

4 &
$-$0.&0621 (21)  &
   0.&00889  (2) &
$-$0.&0532 (21) \\

6 &
$-$0.&0936 (60)  &
   0.&00879  (5) &
$-$0.&0848 (60) \\

8 &
$-$0.&126 (12)   &
   0.&0088  (1)  &
$-$0.&117 (12)  \\

10 &
$-$0.&158 (21)  &
   0.&0087  (1) &
$-$0.&150 (21) \\

12 &
$-$0.&192 (32)  &
   0.&0087  (2) &
$-$0.&183 (32)  \\

14 &
$-$0.&227 (46)  &
   0.&0087  (3) &
$-$0.&218 (46) \\

16 &
$-$0.&263 (62)  &
   0.&0087  (4) &
$-$0.&254 (62) \\

18 &
$-$0.&301 (81)  &
   0.&0087  (5) &
$-$0.&292 (81) \\

20 &
$-$0.&34 (10)   &
   0.&0087  (6) &
$-$0.&33 (10)  \\

24 &
$-$0.&43 (15)  &
   0.&009  (1) &
$-$0.&42 (15) \\

32 &
$-$0.&63 (27)  &
   0.&009  (1) &
$-$0.&62 (27) \\

54 &
$-$1.&6 (8)   &
   0.&009   (4) &
$-$1.&6 (8)  \\

82 &
$-$5.6& (2.0)  &
   0.&009   (9) &
$-$5.6& (2.0) \\

92 &
$-$9.2& (2.6)  &
   0.&009  (12) &
$-$9.2& (2.6) \\

\hline
\hline
\end{tabular}
\end{table}


\squeezetable
\begin{table} \caption{Individual contributions to the
ground-state $g$ factor of Li-like ions.\label{tab:total}} \vspace{0.5cm}
\begin{tabular}{lr@{}lr@{}lr@{}lr@{}l}
\hline
\hline
& \multicolumn{2}{c}{$^{12}{\rm C}^{3+}$} & \multicolumn{2}{c}{$^{16}{\rm O}^{5+}$}
                     & \multicolumn{2}{c}{$^{20}{\rm Ne}^{7+}$} & \multicolumn{2}{c}{$^{24}{\rm Mg}^{9+}$} \\
\hline
Dirac value (point nucleus)&   1.&999 680 300     &     1.&999 431 380       &     1.&999 110 996     &     1.&998 718 893\\
Finite nuclear size        &   0.&000 000 000     &     0.&000 000 000       &     0.&000 000 001     &     0.&000 000 001\\
Interelectronic interaction&   0.&000 130 758 (19)&     0.&000 176 658 (30)  &     0.&000 222 628 (42)&     0.&000 268 703 (55)\\
QED, $\sim \alpha$         &   0.&002 323 017 (1) &     0.&002 323 182 (1)   &     0.&002 323 405 (2) &     0.&002 323 691 (2)\\
QED, $\sim \alpha^2$       &$-$0.&000 003 515     &  $-$0.&000 003 515       &  $-$0.&000 003 516     &  $-$0.&000 003 516\\
Screened QED               &$-$0.&000 000 085 (6) &  $-$0.&000 000 117 (12)  &  $-$0.&000 000 150 (21)&  $-$0.&000 000 183 (32)\\
Nuclear recoil             &   0.&000 000 010     &     0.&000 000 017       &     0.&000 000 025     &     0.&000 000 032\\
Total                      &   2.&002 130 485 (19)&     2.&001 927 604 (32)  &     2.&001 653 389 (47)&     2.&001 307 619 (64)\\
\hline \hline
& \multicolumn{2}{c}{$^{32}{\rm S}^{13+}$}  & \multicolumn{2}{c}{$^{40}{\rm Ar}^{15+}$}
                     & \multicolumn{2}{c}{$^{40}{\rm Ca}^{17+}$}  & \multicolumn{2}{c}{$^{52}{\rm Cr}^{21+}$}  \\
\hline
Dirac value (point nucleus)&     1.&997 718 193    &     1.&997 108 781      &     1.&996 426 011    &     1.&994 838 064\\
Finite nuclear size        &     0.&000 000 005    &     0.&000 000 009      &     0.&000 000 014    &     0.&000 000 035\\
Interelectronic interaction&     0.&000 361 24 (9) &     0.&000 407 75 (12)  &     0.&000 454 45 (14)&     0.&000 548 48 (21)\\
QED, $\sim \alpha$         &     0.&002 324 470 (3)&     0.&002 324 973 (5)  &     0.&002 325 555 (5)&     0.&002 326 984 (5)\\
QED, $\sim \alpha^2$       &  $-$0.&000 003 516 (1)&  $-$0.&000 003 517 (1)  &  $-$0.&000 003 517 (2)&  $-$0.&000 003 518 (6)\\
Screened QED               &  $-$0.&000 000 25 (6) &  $-$0.&000 000 29 (8)   &  $-$0.&000 000 33 (10)&  $-$0.&000 000 42 (15)\\
Nuclear recoil             &     0.&000 000 046 (1)&     0.&000 000 048 (1)  &     0.&000 000 061 (2)&     0.&000 000 070 (4)\\
Total                      &     2.&000 400 19 (11)&     1.&999 837 75 (14)  &     1.&999 202 24 (17)&     1.&997 709 70 (26)\\
\hline
\hline
& \multicolumn{2}{c}{$^{74}{\rm Ge}^{29+}$}   & \multicolumn{2}{c}{$^{132}{\rm Xe}^{51+}$}
                         & \multicolumn{2}{c}{$^{208}{\rm Pb}^{79+}$}  & \multicolumn{2}{c}{$^{238}{\rm U}^{89+}$}   \\
\hline
Dirac value (point nucleus)&     1.&990 752 307     &     1.&972 750 205      &     1.&932 002 904    &     1.&910 722 624 (1)\\
Finite nuclear size        &     0.&000 000 162     &     0.&000 003 37 (1)   &     0.&000 078 64 (16)&     0.&000 241 83 (47)\\
Interelectronic interaction&     0.&000 739 75 (37) &     0.&001 299 4 (11)   &     0.&002 140 7 (27) &     0.&002 501 4 (38)\\
QED, $\sim \alpha$         &     0.&002 330 981 (7) &     0.&002 351 95 (8)   &     0.&002 412 1 (8)  &     0.&002 447 2 (15)\\
QED, $\sim \alpha^2$       &  $-$0.&000 003 523 (24)&  $-$0.&000 003 54 (13)  &  $-$0.&000 003 6 (5)  &  $-$0.&000 003 6 (8)\\
Screened QED               &  $-$0.&000 000 62 (27) &  $-$0.&000 001 6 (8)   &  $-$0.&000 005 6 (20)&  $-$0.&000 009 2 (26)\\
Nuclear recoil             &     0.&000 000 092 (9) &     0.&000 000 16 (6)   &     0.&000 000 25 (35)&     0.&000 000 28 (69)\\
Nuclear polarization       &       &                &       &                 &  $-$0.&000 000 04 (2) &  $-$0.&000 000 27 (14)\\
Total                      &     1.&993 819 14 (46) &     1.&976 399 9 (14)   &     1.&936 625 3(35)&     1.&915 900 2(50)\\
\hline \hline
\end{tabular}
\end{table}

\end{document}